\documentclass[10pt,conference]{IEEEtran}

\usepackage[T1]{fontenc}
\usepackage{amsmath,amssymb,amsthm}
\usepackage{graphicx}
\usepackage{cite}
\usepackage{xcolor}

\newtheorem{theorem}{Theorem}
\newtheorem{proposition}{Proposition}
\newtheorem{corollary}{Corollary}
\newtheorem{remark}{Remark}
\newtheorem{example}{Example}

\newcommand{\Ber}{\operatorname{Ber}}
\newcommand{\E}{\mathbb{E}}
\newcommand{\1}{\mathbf{1}}
\newcommand{\dd}{\mathrm{d}}
\begin{document}

\title{Distribution-Dependent Improvements of McDiarmid's Inequality
via R\'enyi Divergence}

\author{\IEEEauthorblockN{Amedeo Roberto Esposito}
\IEEEauthorblockA{Okinawa Institute of Science and Technology\\
Okinawa, Japan} \and
\IEEEauthorblockN{Adrien Vandenbroucque, Michael Gastpar}
\IEEEauthorblockA{EPFL\\
Lausanne, Switzerland}}

\maketitle

\begin{abstract}
   We investigate whether R\'enyi divergence can sharpen McDiarmid's inequality and its distribution-dependent refinement induced by KL divergence. Starting from Marton's transportation proof, we show that its transport inequality fails in general when KL divergence is replaced by R\'enyi divergence of order $0<\alpha<1$. For sums of independent bounded random variables, however, the additivity of R\'enyi divergence allows us to adapt the argument and derive concentration bounds. For Bernoulli sums, these bounds strictly improve on the distribution-dependent bound induced by KL divergence over a range of deviation thresholds. We also establish a distribution-dependent refinement of Gilardoni's inequality relating total variation to R\'enyi divergence, providing explicit constants for our general concentration bounds.
\end{abstract}


\section{Introduction}\noindent
    Let $X^n=(X_1,\ldots,X_n)$ be a vector of independent random variables. McDiarmid's inequality controls the probability that $f(X^n)$ deviates from its mean when changing the $i$th coordinate changes $f$ by at most $c_i$~\cite{mcdiarmid}. The resulting bound depends on the constants $c_i$, but not on the laws of the $X_i$. Its transportation proof combines Marton's coupling construction with Pinsker's inequality and the chain rule for KL~\cite{marton,raginsky-sason,blm}.
    
    Taking the distribution of $X^n$ into account can yield sharper bounds. Ordentlich and Weinberger refined Pinsker's inequality to take the underlying probability distribution into account~\cite{ordentlich-weinberger}. For biased Bernoulli product measures, their refinement gives a concentration bound with the improved Kearns--Saul constant~\cite{kearns-saul,raginsky-sason}. We compare against this distribution-dependent bound and ask whether using R\'enyi divergence can improve the result further.
    

    This question is motivated by extensions of the connection between transport inequalities and concentration beyond KL divergence~\cite{bobkov-gotze,esposito-gastpar}. We begin by asking whether Marton's transport inequality itself extends to R\'enyi divergence. For every order $0<\alpha<1$, a binary counterexample shows that replacing KL divergence with R\'enyi divergence in this inequality fails in general. For additive functions of independent random variables, however, it suffices to establish a transport inequality between product measures, under which the coordinates remain independent. We obtain such an inequality using the additivity of R\'enyi divergence under products.

    We also prove a distribution-dependent refinement of Gilardoni's inequality \cite{gilardoni}, which gives explicit bounds on the constants in our transport inequality. Finally, we use the inequality to derive concentration results. For sums of Bernoulli random variables, we prove that the bounds obtained using R\'enyi divergence are strictly sharper than the distribution-dependent bound obtained from KL divergence over a range of deviation thresholds. We also briefly mention a continuous example in which this method yields an improvement.

\section{Preliminaries}\noindent
    \subsection{Notation and divergences}\noindent
        All logarithms are natural. For a probability measure $\eta$ and an integrable function $g$, write $\E_\eta[g]$ for the expectation of $g$ under $\eta$.
    
        For probability measures $\nu\ll\mu$, the KL divergence and the R\'enyi divergence of order $0<\alpha<1$ are
        \begin{align}
            D(\nu\Vert\mu)
            &=\int\log\left(\frac{\dd\nu}{\dd\mu}\right)\dd\nu,
            \label{eq:relative-entropy}\\
            D_\alpha(\nu\Vert\mu)
            &=\frac{1}{\alpha-1}
            \log\int\left(\frac{\dd\nu}{\dd\mu}\right)^\alpha \dd\mu.
            \label{eq:renyi}
        \end{align}
        For probability measures $\rho$ and $\sigma$ on the same measurable space, their total variation distance is $\|\rho-\sigma\|_{\rm TV}=\sup_A|\rho(A)-\sigma(A)|$, where the supremum is over measurable events $A$.
    
    \subsection{McDiarmid's inequality via KL}\noindent
    Let $X^n$ have independent coordinates and joint law $\mu=\mu_1\otimes\cdots\otimes\mu_n$, and let $f:\mathcal X^n\to\mathbb R$ satisfy
    \begin{equation}
        |f(x^n)-f(y^n)|\leq
        \sum_{i=1}^n c_i\1_{\{x_i\neq y_i\}}.
        \label{eq:bounded-differences}
    \end{equation}
    Set $C=\sum_{i=1}^n c_i^2$. If $C=0$, then $f$ is constant. For $C>0$, McDiarmid's inequality states that, for every $t>0$,
    \begin{equation}
        \mu\bigl(f-\E_\mu[f]\geq t\bigr)\leq e^{-2t^2/C}.
        \label{eq:mcdiarmid}
    \end{equation}
    We recall its proof via transportation-cost inequalities~\cite[Section 3.4.3]{raginsky-sason-arxiv}. For $\nu\ll\mu$, let $\Pi(\mu,\nu)$ denote the set of joint laws of $(X^n,Y^n)$ with marginals $\mu$ and $\nu$. Define Marton's cost~\cite{marton} by
    \begin{equation}
        \mathcal M^2(\nu,\mu)
        =\inf_{\pi\in\Pi(\mu,\nu)}
        \sum_{i=1}^n\pi(X_i\neq Y_i)^2.
        \label{eq:marton-cost}
    \end{equation}
    
    Bounded differences and Cauchy--Schwarz give
    \begin{align}
        \big|\E_\nu[f]-\E_\mu[f]\big|
        &\leq\inf_{\pi\in\Pi(\mu,\nu)}
        \sum_{i=1}^n c_i\pi(X_i\neq Y_i) \notag\\
        &\leq\sqrt{C\mathcal M^2(\nu,\mu)}.
    \end{align}
    To bound this cost, Marton's construction couples the coordinates sequentially. Given $Y^{i-1}=y^{i-1}$, it uses a maximal coupling of $\mu_i$ and the conditional law $\nu_{i\mid y^{i-1}}$ of $Y_i$, minimising the probability that $X_i\neq Y_i$. This probability is their total variation distance. The resulting coupling satisfies
    \begin{equation}
        \pi(X_i\neq Y_i)
        =\int\|\nu_{i\mid y^{i-1}}-\mu_i\|_{\rm TV}
        \,\dd\nu^{i-1}(y^{i-1}).
    \end{equation}
    Applying Pinsker's inequality to each conditional law and then Jensen's inequality gives
    \begin{align}
        \mathcal M^2(\nu,\mu)
        &\leq\sum_{i=1}^n \pi(X_i\neq Y_i)^2
        \notag\\
        &\leq \frac12\sum_{i=1}^n
        \!\int\! D(\nu_{i\mid y^{i-1}}\Vert\mu_i)
        \,\dd\nu^{i-1}(y^{i-1}) \!=\!\frac12D(\nu\Vert\mu).
        \label{eq:marton-kl}
    \end{align}
    The last equality is the chain rule for KL.
    
    Combining the two estimates gives
    \begin{equation}
        \big|\E_\nu[f]-\E_\mu[f]\big|
        \leq\sqrt{\frac{C}{2}D(\nu\Vert\mu)}.
        \label{eq:kl-transport}
    \end{equation}
    The Donsker--Varadhan representation of KL divergence~\cite{donsker-varadhan} now bounds the exponential moments: for $\lambda>0$,
    \begin{align}
        \log\E_\mu\Bigl[e^{\lambda(f-\E_\mu[f])}\Bigr]
        &=\sup_{\nu\ll\mu}
        \Big\{\lambda\big(\E_\nu[f]-\E_\mu[f]\big)-D(\nu\Vert\mu)\Big\}\notag\\
        &\leq\sup_{d\geq0}\left\{\lambda\sqrt{\frac{Cd}{2}}-d\right\}
        =\frac{C\lambda^2}{8}.
        \label{eq:mcdiarmid-mgf}
    \end{align}
    Exponential Markov's inequality gives
    \[
        \mu\bigl(f-\E_\mu[f]\geq t\bigr)
        \leq e^{-\lambda t+\frac{C\lambda^2}{8}}.
    \]
    Choosing $\lambda=4t/C$ proves \eqref{eq:mcdiarmid}.

\section{An Obstruction for Arbitrary Measures}\noindent
    The preceding proof uses the chain rule to combine the conditional KL divergences into $D(\nu\Vert\mu)$. R\'enyi divergence does not satisfy this chain rule. More strongly, replacing KL by R\'enyi divergence in \eqref{eq:marton-kl} fails even when the reference measure is a product. The counterexample below changes every marginal while retaining a common component with the reference law. As the dimension increases, its transport cost grows, whereas its R\'enyi divergence remains bounded.
    \begin{theorem}\label{thm:impossibility}
        Fix $0<\alpha<1$. There is no finite constant $A_\alpha$, independent
        of $n$, such that
        \begin{equation}
            \mathcal M^2(\nu,\mu)
            \leq A_\alpha D_\alpha(\nu\Vert\mu)
            \label{eq:false-marton}
        \end{equation}
        for every product measure $\mu$ and every $\nu\ll\mu$.
    \end{theorem}
    
    \begin{proof}
        Fix $s\in(0,1)$ and let
        \begin{equation}
            \mu=\Ber(1/2)^{\otimes n},\qquad
            \nu=(1-s)\mu+s\delta_{1^n}.
            \label{eq:counterexample}
        \end{equation}
        The $i$th marginal of $\nu$ is $\Ber((1+s)/2)$. Hence, for every coupling $\pi\in\Pi(\mu,\nu)$,
        \begin{equation}
            \pi(X_i\neq Y_i)
            \geq \left\|\mu_i-\nu_i\right\|_{\rm TV}
            =\frac{s}{2}.
            \label{eq:marginal-lower}
        \end{equation}
        It follows that $\mathcal M^2(\nu,\mu)\geq ns^2/4$.
        
        This lower bound is attained. Draw $X^n\sim\mu$ and
        $B\sim\Ber(s)$ independently, and set $Y^n=X^n$ if $B=0$ and
        $Y^n=1^n$ if $B=1$. Then $Y^n\sim\nu$ and
        \begin{equation}
            \pi(X_i\neq Y_i)=\Pr(B=1,X_i=0)=\frac{s}{2}.
        \end{equation}
        Consequently,
        \begin{equation}
            \mathcal M^2(\nu,\mu)=\frac{ns^2}{4}.
            \label{eq:counter-cost}
        \end{equation}
        
        The density $\dd\nu/\dd\mu$ equals $1-s$ away from $1^n$ and $1-s+s2^n$ at $1^n$. Substitution in \eqref{eq:renyi} gives
        $D_\alpha(\nu\Vert\mu)=\frac{1}{\alpha-1}\log\bigl[       (1-2^{-n})(1-s)^\alpha
            +2^{-n}(1-s+s2^n)^\alpha\bigr].$
        Since $0<\alpha<1$, the second term in the brackets is of order $2^{-n(1-\alpha)}$. Therefore
        \begin{equation}
            \lim_{n\to\infty}D_\alpha(\nu\Vert\mu)
            =-\frac{\alpha}{1-\alpha}\log(1-s)<\infty,
            \label{eq:counter-limit}
        \end{equation}
        and hence equations~\eqref{eq:counter-cost} and~\eqref{eq:counter-limit} rule out any constant independent of $n$. For the same pair, $D(\nu\Vert\mu)=sn\log 2+O(1)$, consistent with \eqref{eq:marton-kl}.
    \end{proof}

\section{Additive Functions}\noindent
    The obstruction above concerns arbitrary measures $\nu$. For product measures, R\'enyi divergence is additive, which allows us to combine estimates across coordinates. We therefore consider functions of the form
    \begin{equation}
        f(x^n)=\sum_{i=1}^n f_i(x_i),
        \label{eq:additive}
    \end{equation}
    where each $f_i$ is bounded. For $\rho=\bigotimes_i\rho_i$ and $\mu=\bigotimes_i\mu_i$, the difference of expectations is the sum of the coordinate differences $\E_{\rho_i}[f_i]-\E_{\mu_i}[f_i]$. For $0<\alpha<1$, let
        \begin{equation}
            \kappa_{\alpha,i}
            =\sup_{\substack{\rho_i\ll\mu_i\\\rho_i\neq\mu_i}}
            \frac{\big(\E_{\rho_i}[f_i]-\E_{\mu_i}[f_i]\big)^2}
            {2D_\alpha(\rho_i\Vert\mu_i)},
            \qquad C_\alpha=\sum_{i=1}^n\kappa_{\alpha,i}.
            \label{eq:kappa}
        \end{equation}
        Thus $\kappa_{\alpha,i}$ is the smallest constant for which
        \begin{equation}
            \big|\E_{\rho_i}[f_i]-\E_{\mu_i}[f_i]\big|
            \leq\sqrt{2\kappa_{\alpha,i}D_\alpha(\rho_i\Vert\mu_i)}
            \label{eq:coordinate-bound}
        \end{equation}
        holds for every $\rho_i\ll\mu_i$. For a constant summand, we set $\kappa_{\alpha,i}=0$. The following result combines these inequalities using Cauchy--Schwarz and the additivity of R\'enyi divergence.
        
        \begin{proposition}
            \label{prop:boundedDifferencesOfMeans}
            \label{thm:product-transport}
            Let $\mu=\bigotimes_{i=1}^n\mu_i$ and $f(x^n)=\sum_{i=1}^n f_i(x_i)$, with bounded $f_i$ and $C_\alpha$ defined in \eqref{eq:kappa}. Let $\alpha \in (0,1)$. For every measure $\rho=\bigotimes_{i=1}^n\rho_i$ with $\rho_i\ll\mu_i$ for every $i$,
            \begin{equation}
                \big|\E_\rho[f]-\E_\mu[f]\big|
                \leq\sqrt{2C_\alpha D_\alpha(\rho\Vert\mu)}.
                \label{eq:product-transport}
            \end{equation}
        \end{proposition}
        \begin{remark}
For additive functions, the boundedness assumption corresponds
to the bounded-differences condition in McDiarmid's inequality.
Marton's method has the same boundedness requirement in the
additive setting, since bounded differences then require each
summand to have bounded oscillation. However, the corresponding bound
 applies to every $\nu\ll\mu$,
whereas Proposition~\ref{prop:boundedDifferencesOfMeans} requires both measures
to be product measures.
\end{remark}

    \subsection{A distribution-dependent version of Gilardoni's inequality}\noindent
        Theorem~\ref{thm:product-transport} uses the optimal coordinate constants, but computing them requires an optimisation over probability measures. Its proof also applies if each $\kappa_{\alpha,i}$ is replaced by an upper bound. We now derive such a bound, and in doing so, we obtain a refinement of Gilardoni's inequality~\cite[Corollary 9]{gilardoni}, which states that for $0<\alpha<1$,
        \begin{equation}\label{eq:gilardoni}
            D_\alpha(\nu\Vert\mu)
            \geq 2\alpha\|\nu-\mu\|_{\mathrm{TV}}^2.
        \end{equation}
        
        We first state the result for bounded functions so that it applies directly to the coordinate differences in \eqref{eq:coordinate-bound}.
        
        Define $
            \psi(r):=\frac{1-2r}{\log((1-r)/r)},
            \,\, 0<r<1,$
        with continuous values $\psi(0)=\psi(1)=0$ and $\psi(1/2)=1/2$. In particular, $0\leq\psi(r)\leq1/2$.
        
        \begin{proposition}\label{prop:refined-pinsker}
            Let $0<\alpha<1$, let $\nu\ll\mu$, and let $g$ be a measurable
            function taking values in $[0,1]$. Then
            \begin{align}
                &\bigl(\E_\nu[g]\!-\!\E_\mu[g]\bigr)^2\!\!\leq\!\!
                \Bigl[
                \alpha\psi\bigl(\E_\mu[g]\bigr)
                \!+\!(1-\alpha)\psi\bigl(\E_\nu[g]\bigr)
                \Bigr]\frac{D_\alpha(\nu\Vert\mu)}{\alpha}.
                \label{eq:refined-expectation}
            \end{align}
        \end{proposition}
        \begin{proof}
            The refined Hoeffding bound~\cite[Section 2.2.4]{raginsky-sason} gives, for any probability measure $\eta$ and any $s\in\mathbb R$,
            \begin{equation}
                \log\E_\eta\!\left[e^{s\big(g-\E_\eta[g]\big)}\right]
                \leq \frac{s^2}{4}\psi\bigl(\E_\eta[g]\bigr).
                \label{eq:refined-hoeffding}
            \end{equation}
            For every $\lambda\in\mathbb R$, the variational representation of R\'enyi divergence (see e.g.~\cite[Lemma 2]{esposito-gastpar-issa}) yields
            \begin{align}
                \frac{D_\alpha(\nu\Vert\mu)}{\alpha}
                &\geq
                \frac{1}{\alpha-1}
                \log\E_\nu\!\left[e^{(\alpha-1)\lambda g}\right]
                -
                \frac{1}{\alpha}
                \log\E_\mu\!\left[e^{\alpha\lambda g}\right].
                \label{eq:renyi-variational-test}
            \end{align}
            Apply \eqref{eq:refined-hoeffding} to both logarithms. Since $\alpha-1<0$, division by $\alpha-1$ reverses the inequality for the first term. Consequently,
            \begin{align}
                \frac{D_\alpha(\nu\Vert\mu)}{\alpha}
                &\geq
                \lambda\bigl(\E_\nu[g]-\E_\mu[g]\bigr)
                \notag\\
                &\quad-\frac{\lambda^2}{4}
                \Bigl[
                \alpha\psi\bigl(\E_\mu[g]\bigr)
                +(1-\alpha)\psi\bigl(\E_\nu[g]\bigr)
                \Bigr].
            \end{align}
            Maximising the right-hand side over $\lambda\in\mathbb R$ proves \eqref{eq:refined-expectation} whenever the coefficient of $\lambda^2$ is positive. If it vanishes, then $\E_\mu[g]\in\{0,1\}$, so $g$ is constant $\mu$-almost surely. Absolute continuity implies that the two expectations coincide, and the claim follows.
        \end{proof}
        Proposition~\ref{prop:refined-pinsker} naturally leads to the following distribution-dependent version of~\eqref{eq:gilardoni}, thereby extending the distribution-dependent Pinsker's inequality from~\cite[Theorem 2.1]{ordentlich-weinberger}.
        \begin{corollary}\label{cor:refined-gilardoni}
            Let $0<\alpha<1$ and let $\nu\ll\mu$. Then
            \begin{equation}
                \|\nu-\mu\|_{\mathrm{TV}}^2
                \leq
                \Gamma_\alpha(\nu,\mu)
                \frac{D_\alpha(\nu\Vert\mu)}{\alpha},
                \label{eq:refined-gilardoni}
            \end{equation}
            where $
                \Gamma_\alpha(\nu,\mu)
                :=
               \{\alpha\psi\bigl(\mu(A_\star)\bigr)
                +(1-\alpha)\psi\bigl(\nu(A_\star)\bigr)
                \}$, where $A_\star = \{d\nu/d\mu \geq 1\}.$
        \end{corollary}
        \begin{proof}
           Take $g=\mathbf1_{A_\star}$ in Proposition~\ref{prop:refined-pinsker}.
        \end{proof}
        If $\nu \neq \mu$ one has that $\Gamma_\alpha(\nu,\mu)<1/2$, yielding an improvement over Gilardoni's quadratic result.
        
        Suppose now that $f_i$ takes values in $[a_i,b_i]$. Applying Proposition~\ref{prop:refined-pinsker} to $(f_i-a_i)/(b_i-a_i)$ and bounding the term involving $\rho_i$ by $\psi\leq1/2$ gives
        \begin{equation}
            \kappa_{\alpha,i}\leq\frac{(b_i-a_i)^2}{2\alpha}
            \left[
            \alpha\psi\!\left(\frac{\E_{\mu_i}[f_i]-a_i}{b_i-a_i}\right)
            +\frac{1-\alpha}{2}
            \right].
            \label{eq:explicit-coordinate-constant}
        \end{equation}
        The right-hand side depends only on the range and expectation of $f_i$ under $\mu_i$. Substituting these upper bounds into Theorem~\ref{thm:product-transport} therefore gives an estimate that requires no optimisation over the measures $\rho_i$.
    
    \subsection{Concentration via R\'enyi divergence}\noindent
        To obtain a tail bound from Theorem \ref{thm:product-transport}, we bound the exponential moments of $f$. Replace each $f_i$ by $f_i-\E_{\mu_i}[f_i]$, so that their sum $f$ has mean zero, and define the logarithmic moment-generating function (log-MGF)
        \begin{equation}
            K(\lambda):=\log\E_\mu\left[e^{\lambda f}\right].
            \label{eq:cgf}
        \end{equation}
        For $\lambda>0$, the variational supremum in the first line of \eqref{eq:mcdiarmid-mgf} is attained at the probability measure $\nu_\lambda$ defined through
        \begin{equation}
            \frac{\dd\nu_\lambda}{\dd\mu}
            :=e^{\lambda f-K(\lambda)}.
            \label{eq:tilt}
        \end{equation}
        This measure is called the exponential tilt of $\mu$ by $\lambda f$. It suffices to control the difference of expectations for this measure, since it attains the variational supremum. Since $f$ is additive, the normalising constant in \eqref{eq:tilt} factors: $\E_\mu[e^{\lambda f}]=\prod_i\E_{\mu_i}[e^{\lambda f_i}]$. Consequently, $\nu_\lambda=\bigotimes_i\nu_{\lambda,i}$, where $\dd\nu_{\lambda,i}/\dd\mu_i=e^{\lambda f_i}/\E_{\mu_i}[e^{\lambda f_i}]$. We can therefore apply Theorem \ref{thm:product-transport} to $\nu_\lambda$. To compare with the bound obtained from KL divergence, define
        \begin{equation}
            E_\alpha(\lambda)
            =D(\nu_\lambda\Vert\mu)-D_\alpha(\nu_\lambda\Vert\mu)
            \geq0.
            \label{eq:correction-divergence}
        \end{equation}
        The inequality follows from monotonicity of R\'enyi divergence in its order~\cite[Theorem 3]{vanerven}.
        
        \begin{theorem}\label{thm:mgf}
            Under the assumptions of Theorem \ref{thm:product-transport}, with $\E_\mu[f]=0$, for every $\lambda>0$,
            \begin{equation}
                K(\lambda)
                \leq\frac{C_\alpha\lambda^2}{2}-E_\alpha(\lambda).
                \label{eq:mgf-bound}
            \end{equation}
            For every $t>0$,
            \begin{equation}
                \mu(f\geq t)
                \leq\inf_{\lambda>0}e^{-\lambda t
                +\frac{C_\alpha\lambda^2}{2}-E_\alpha(\lambda)}.
                \label{eq:tail-bound}
            \end{equation}
        \end{theorem}
        
        \begin{proof}
            If $C_\alpha=0$, then $f$ is $\mu$-almost surely constant and the centred claim is immediate. Assume $C_\alpha>0$. Since $f$ is additive, $\nu_\lambda$ is a product measure. Differentiating \eqref{eq:cgf} gives
            \begin{equation}
                \E_{\nu_\lambda}[f]=K'(\lambda).
                \label{eq:tilt-mean}
            \end{equation}
            Substituting \eqref{eq:tilt} into the definition of R\'enyi divergence gives
            \begin{equation}
                D_\alpha(\nu_\lambda\Vert\mu)
                =\frac{\alpha K(\lambda)-K(\alpha\lambda)}{1-\alpha},
                \label{eq:tilt-renyi}
            \end{equation}
            whereas for KL divergence one has
            \begin{equation}
                D(\nu_\lambda\Vert\mu)
                =\lambda K'(\lambda)-K(\lambda).
                \label{eq:tilt-kl}
            \end{equation}
            Subtracting these identities characterises the correction in terms of $K$:
            \begin{equation}
                E_\alpha(\lambda)
                =\lambda K'(\lambda)
                -\frac{K(\lambda)-K(\alpha\lambda)}{1-\alpha}.
                \label{eq:correction}
            \end{equation}
            
            Theorem \ref{thm:product-transport} and \eqref{eq:tilt-mean} imply
            \begin{equation}
                K'(\lambda)^2
                \leq2C_\alpha D_\alpha(\nu_\lambda\Vert\mu).
                \label{eq:mean-divergence}
            \end{equation}
            By \eqref{eq:correction-divergence} and \eqref{eq:tilt-kl},
            \begin{equation}
                D_\alpha(\nu_\lambda\Vert\mu)
                =\lambda K'(\lambda)-K(\lambda)-E_\alpha(\lambda).
            \end{equation}
            Substitution in \eqref{eq:mean-divergence} and completion of the square give
            \begin{align}
                K(\lambda)
                &\leq\lambda K'(\lambda)
                -\frac{K'(\lambda)^2}{2C_\alpha}-E_\alpha(\lambda)\\
                &\leq\frac{C_\alpha\lambda^2}{2}-E_\alpha(\lambda).
                \label{eq:complete-square}
            \end{align}
            Applying Markov's inequality to $e^{\lambda f}$ gives
            \begin{equation}
                \mu\bigl(f-\E_\mu[f]\geq t\bigr)
                =\mu(f\geq t)\leq e^{-\lambda t+K(\lambda)}.
            \end{equation}
            Applying \eqref{eq:mgf-bound} and taking the infimum over $\lambda>0$ proves \eqref{eq:tail-bound}.
        \end{proof}
        
        \begin{remark}
            For a general function satisfying bounded differences, the measure \eqref{eq:tilt} need not be a product. In fact, the measure $\nu$ in \eqref{eq:counterexample} has this form for $g(x^n)=\1_{\{x^n=1^n\}}-2^{-n}$ and $\lambda=\log[(1-s+s2^n)/(1-s)]$. Changing one coordinate changes $g$ by at most one. Thus the obstruction in Theorem \ref{thm:impossibility} also occurs among exponential tilts of functions satisfying bounded differences.
        \end{remark}
        
        Define $\kappa_{1,i}$ by \eqref{eq:kappa}, replacing $D_\alpha$ with KL divergence, and set $C_1=\sum_i\kappa_{1,i}$. Repeating the proof with KL divergence throughout yields $
            K(\lambda)\leq \frac{C_1\lambda^2}{2}.  $
        This represents our main source of comparison for the bounds we introduce. Since $D_\alpha\leq D$ for $0<\alpha<1$, the constants satisfy $\kappa_{\alpha,i}\geq\kappa_{1,i}$ and hence $C_\alpha\geq C_1$. The quadratic term in the bound using R\'enyi divergence is therefore at least as large as its KL divergence counterpart and, on its own, gives no improvement. Any gain must come from the correction term: at a given $\lambda$, the amount it subtracts must exceed
        \begin{equation}
            E_\alpha(\lambda)>\frac{(C_\alpha-C_1)\lambda^2}{2}.
            \label{eq:improvement-condition}
        \end{equation}
        Thus, the comparison depends on the balance between the larger quadratic coefficient and the correction, rather than on the ordering of the divergences alone.
        
        If $c_i=\sup f_i-\inf f_i$, Pinsker's inequality gives $\kappa_{1,i}\leq c_i^2/4$ and recovers McDiarmid's inequality. 
        Since $K''(u)=\operatorname{Var}_{\nu_u}(f)$, setting
$v_\lambda=\inf_{\alpha\lambda\leq u\leq\lambda}K''(u)$ gives
\begin{align}
E_\alpha(\lambda)
&=\frac{1}{1-\alpha}\int_{\alpha\lambda}^{\lambda}
(u-\alpha\lambda)K''(u)\,\dd u \notag\\
&\geq \frac{(1-\alpha)v_\lambda\lambda^2}{2}.
\label{eq:variance-correction}
\end{align}
Consequently, $K(\lambda)\leq
\big(C_\alpha-(1-\alpha)v_\lambda\big)\lambda^2/2$,
which improves on the quadratic KL divergence bound whenever
$(1-\alpha)v_\lambda>C_\alpha-C_1$.
\begin{figure*}[!t]
         \centering
         \includegraphics[width=0.93\textwidth]{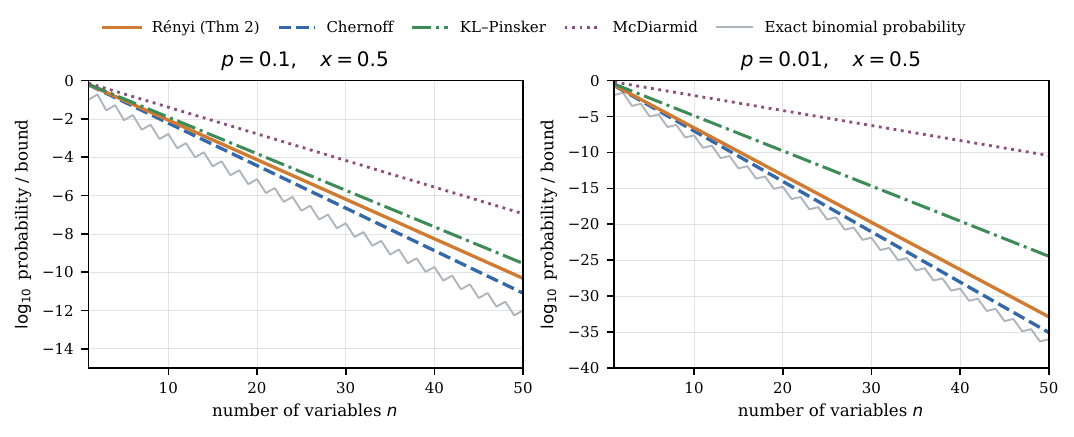}
     \caption{Bounds on     $\Pr(\overline X_n\geq0.5)$ for
    (a) $p=0.1$ and (b) $p=0.01$.
    The four bounds are plotted as $\log_{10}B_n$, where
    $B_n=\exp(-nI)$. The exponent from Theorem~\ref{thm:mgf}
    is computed from \eqref{eq:binary-rate}, optimising
    $\alpha$ and $\theta$ numerically. Writing $u=0.5-p$,
    the other exponents are $u^2/(2\kappa_1(p))$ for
    KL--Pinsker, $2u^2$ for McDiarmid, and
    $D(\mathrm{Bern}(0.5)\Vert\mathrm{Bern}(p))$ for Chernoff. The optimising pairs $(\alpha,\theta)$ are approximately $(0.83,2.36)$ for $p=0.1$ and $(0.77,5.24)$ for $p=0.01$. The thin grey curve shows the exact binomial probability.}
\label{fig:binary}
\end{figure*}
\section{Bernoulli Sums}\label{sec:bernoulliSums}\noindent
    Let $X_1,\ldots,X_n$ be i.i.d. $\Ber(p)$, with $0<p<1$, and set $\overline X_n=n^{-1}\sum_{i=1}^nX_i$. For the centred sum $f=\sum_{i=1}^n(X_i-p)$, independence gives
    \begin{equation}
        K(\lambda)=nk_p(\lambda),\qquad k_p(\lambda)=\log(1-p+pe^\lambda)-p\lambda.
        \label{eq:binary-cgf}
    \end{equation}
    Since the log-MGF is known in this setting, the Chernoff bound can be computed exactly and yields
    \begin{equation}
        \Pr\bigl(\overline X_n-p\geq u\bigr)
        \leq e^{-n\sup_{\lambda>0}\{\lambda u-k_p(\lambda)\}}=e^{-nd(p+u\Vert p)},
        \label{eq:binary-chernoff}
    \end{equation}
    where $0<u<1-p$ and $d(q\Vert p)$ denotes the KL divergence between the Bernoulli laws with parameters $q$ and $p$.

    In comparison, the estimate $k_p(\lambda)\leq\lambda^2/8$ gives McDiarmid's bound $e^{-2nu^2}$. Moreover, the distribution-dependent bound obtained from KL divergence~\cite{ordentlich-weinberger,raginsky-sason,kearns-saul} improves this estimate to $k_p(\lambda)\leq\kappa_1(p)\lambda^2/2$, where $\kappa_1(p)=\psi(p)/2$. The resulting tail bound is $e^{-nI_D(u)}$, where
    \begin{equation}
        I_D(u)=\frac{u^2}{2\kappa_1(p)}.
        \label{eq:kearns-saul}
    \end{equation}

    Let us now compute the bound from Theorem~\ref{thm:mgf}.
   The coordinate constant becomes
    \begin{equation}
        \kappa_\alpha(p)
        =\sup_{\substack{q\in[0,1]\\q\neq p}}\frac{(q-p)^2}{2d_\alpha(q\Vert p)},
        \qquad C_\alpha=n\kappa_\alpha(p),
        \label{eq:binary-kappa}
    \end{equation}
    where $d_\alpha(q\Vert p)$ is the binary R\'enyi divergence between the Bernoulli laws with parameters $q$ and $p$.
    Under the tilted law $\nu_{\lambda,i}$, the Bernoulli parameter is $
        q_\lambda=pe^\lambda/(1-p+pe^\lambda).
        \label{eq:binary-tilt}$
    Substituting $K=nk_p$ and $K'(\lambda)=n(q_\lambda-p)$ into \eqref{eq:correction} gives
    \begin{equation}
        \frac{E_\alpha(\lambda)}{n}
        =\lambda(q_\lambda-p)
        -\frac{k_p(\lambda)-k_p(\alpha\lambda)}{1-\alpha}.
        \label{eq:binary-correction}
    \end{equation}
    A deviation of the sample mean by $u$ corresponds to $t=nu$ in Theorem~\ref{thm:mgf}. The theorem therefore gives
    \begin{equation}
        \Pr\bigl(\overline X_n-p\geq u\bigr)\leq e^{-nI_{\rm R}(u)},
        \label{eq:binary-tail}
    \end{equation}
    where
    \begin{equation}
        I_{\rm R}(u)
        =\sup_{\substack{0<\alpha<1\\\lambda>0}}
        \left\{\lambda u-\frac{\kappa_\alpha(p)\lambda^2}{2}
        +\frac{E_\alpha(\lambda)}{n}\right\}.
        \label{eq:binary-rate}
    \end{equation}
    By \eqref{eq:binary-correction}, this rate does not depend on $n$. The following proposition establishes a strict improvement over $I_D(u)$.

    \begin{proposition}\label{prop:strict}
        If $0<p<1/2$ and $0<u<1-2p$, then
        \begin{equation}
            I_{\rm R}(u)>I_D(u).
            \label{eq:strict-improvement}
        \end{equation}
    \end{proposition}

    \begin{proof}
        We keep the optimiser defining $I_D(u)$ fixed and show that decreasing $\alpha$ slightly below $1$ increases the rate. Let $L=\log((1-p)/p)$, $d=1-2p$, and $
            \lambda_0=\frac{u}{\kappa_1(p)}=\frac{2Lu}{d}.
            \label{eq:lambda-zero}$
        
        We first compute the change in the optimal coordinate constant. For $\varepsilon\geq0$ close to zero, let $F_\varepsilon(q)=(q-p)^2/[2D_{1-\varepsilon}(q\Vert p)]$, with continuous value $F_\varepsilon(p)=p(1-p)/(1-\varepsilon)$. Here $D_1=D$. The function $F_\varepsilon$ and its derivative with respect to $\varepsilon$ are jointly continuous on a neighbourhood of $\{0\}\times[0,1]$. At $\varepsilon=0$, the unique maximiser is $q=1-p$~\cite{ordentlich-weinberger}. Danskin's theorem~\cite{danskin} therefore shows that the first-order change in the maximum is obtained by keeping $q=1-p$ fixed and differentiating $F_\varepsilon(q)$ with respect to $\varepsilon$. Since
       $
            D_{1-\varepsilon}(1-p\Vert p)
            =dL-2\varepsilon p(1-p)L^2+o(\varepsilon),
        $
        we obtain $
            \kappa_{1-\varepsilon}(p)
            =\kappa_1(p)+\varepsilon p(1-p)+o(\varepsilon).
            \label{eq:kappa-expansion}$
        Taylor expansion of \eqref{eq:binary-correction}, using $k_p''(\lambda)=q_\lambda(1-q_\lambda)$, similarly gives
       $
            \frac{E_{1-\varepsilon}(\lambda)}{n}
            =\frac{\varepsilon\lambda^2}{2}q_\lambda(1-q_\lambda)
            +o(\varepsilon).
            \label{eq:correction-expansion}$
        The first expansion determines the increase in the quadratic penalty; the second gives the gain from the correction. Evaluating the objective in \eqref{eq:binary-rate} at $\alpha=1-\varepsilon$ and $\lambda=\lambda_0$ now yields
        \begin{align}
            I_{\rm R}(u)&\geq I_D(u)
            +\frac{\varepsilon\lambda_0^2}{2}
            \bigl[q_{\lambda_0}(1-q_{\lambda_0})-p(1-p)\bigr]+o(\varepsilon).
            \label{eq:rate-expansion}
        \end{align}
        The assumption $0<u<d$ gives $0<\lambda_0<2L$. Since $q_0=p$, $q_{2L}=1-p$, and $q_\lambda$ is strictly increasing, it follows that $p<q_{\lambda_0}<1-p$. Thus $q_{\lambda_0}(1-q_{\lambda_0})>p(1-p)$, and the rate strictly exceeds $I_D(u)$ for sufficiently small $\varepsilon>0$.
    \end{proof}


    Figure \ref{fig:binary} compares numerically four upper bounds on $\Pr(\overline X_n\geq1/2)$ as the sample size $n$ varies: our bound using R\'enyi divergence, the KL divergence bound, McDiarmid's bound, and the exact Chernoff bound. For comparison, we also plot the exact probability, computed numerically from the binomial distribution. The two plots use $p=0.1$ and $p=0.01$. In both cases, Theorem \ref{thm:mgf} gives a smaller bound than the bound derived from the refinement of Pinsker's inequality~\cite{ordentlich-weinberger} and a larger bound than exact Chernoff.
    \begin{remark}
      A similar approach to the one developed in this paper can also be employed for R\'enyi divergences of order $\alpha>1$, yielding improvements for larger deviations. For instance, in the Bernoulli example with $0<p<1/2$, choosing $\alpha>1$ sufficiently close to $1$ improves the bound induced by KL divergence for the event $\{\overline X_n-p\geq u\}$ when $1-2p<u<1-p$. This complements the improvement obtained with $\alpha<1$ for $0<u<1-2p$. At $u=1-2p$, the bounds obtained using R\'enyi divergence coincide with the bound induced by KL divergence, which
also matches the Chernoff bound.
    \end{remark}
    \begin{example}\label{ex:continuous}
        The improvement is not restricted to Bernoulli sums or discrete distributions. For example, consider independent random variables with the truncated exponential density
        $
        h(x)=\frac{12e^{-12x}}{1-e^{-12}}\1_{[0,1]}(x).
        $
        For the event $\overline X_n\geq1/2$, choosing $\alpha=0.9$ and $\lambda=13$ in Theorem~\ref{thm:mgf}, one can numerically verify that the exponential rate obtained using R\'enyi divergence is approximately $2.92173$, compared with $2.50004$ for the bound obtained from KL divergence. For $n=10$, the resulting upper bound is approximately $68$ times smaller.
    \end{example}
    \newpage

\begin{thebibliography}{99}

\bibitem{mcdiarmid}
C. McDiarmid, ``On the method of bounded differences,'' in
\emph{Surveys in Combinatorics}, London Math. Soc. Lecture Note Ser.,
vol. 141. Cambridge, U.K.: Cambridge Univ. Press, 1989, pp. 148--188.

\bibitem{marton}
K. Marton, ``Bounding $\bar d$-distance by informational divergence: A
method to prove measure concentration,'' \emph{Ann. Probab.}, vol. 24,
no. 2, pp. 857--866, 1996.

\bibitem{ordentlich-weinberger}
E. Ordentlich and M. J. Weinberger, ``A distribution dependent refinement
of Pinsker's inequality,'' \emph{IEEE Trans. Inf. Theory}, vol. 51, no. 5,
pp. 1836--1840, May 2005.

\bibitem{blm}
S. Boucheron, G. Lugosi, and P. Massart, \emph{Concentration
Inequalities: A Nonasymptotic Theory of Independence}. Oxford, U.K.:
Oxford Univ. Press, 2013.

\bibitem{raginsky-sason}
M. Raginsky and I. Sason, ``Concentration of measure inequalities in
information theory, communications, and coding,'' \emph{Found. Trends
Commun. Inf. Theory}, vol. 10, nos. 1--2, pp. 1--246, 2013.

\bibitem{raginsky-sason-arxiv}
M. Raginsky and I. Sason, ``Concentration of Measure Inequalities in Information
Theory, Communications and Coding (Second Edition),'' \emph{arXiv:1212.4663 [cs.IT]},
2015.

\bibitem{vanerven}
T. van Erven and P. Harremo\"es, ``R\'enyi divergence and
Kullback--Leibler divergence,'' \emph{IEEE Trans. Inf. Theory}, vol. 60,
no. 7, pp. 3797--3820, Jul. 2014.

\bibitem{gilardoni}
G. L. Gilardoni, ``On Pinsker's and Vajda's Type Inequalities for Csiszár's $f$ -Divergences,'' in \emph{IEEE Trans. Inf. Theory}, vol. 56, no. 11, pp. 5377-5386, Nov. 2010.

\bibitem{kearns-saul}
M. J. Kearns and L. K. Saul, ``Large deviation methods for approximate
probabilistic inference,'' in \emph{Proc. 14th Conf. Uncertainty Artif.
Intell.}, 1998, pp. 311--319.

\bibitem{esposito-gastpar}
A. R. Esposito and M. Gastpar, ``From generalisation error to
transportation-cost inequalities and back,'' in \emph{Proc. IEEE Int.
Symp. Inf. Theory}, 2022, pp. 294--299.

\bibitem{esposito-gastpar-issa}
A. R. Esposito, M. Gastpar, and I. Issa, ``Sibson $\alpha$-Mutual Information and Its Variational Representations,'' \emph{IEEE Trans. Inf. Theory}, vol. 72, no. 7, pp. 4434--4467, Jul. 2026.

\bibitem{bobkov-gotze}
S. G. Bobkov and F. G\"otze, ``Exponential integrability and
transportation cost related to logarithmic Sobolev inequalities,''
\emph{J. Funct. Anal.}, vol. 163, no. 1, pp. 1--28, 1999.
\bibitem{donsker-varadhan}
M. D. Donsker and S. R. S. Varadhan, ``Asymptotic evaluation of
certain Markov process expectations for large time, I,''
\emph{Commun. Pure Appl. Math.}, vol. 28, no. 1, pp. 1--47,
1975, doi: 10.1002/cpa.3160280102.
\bibitem{rudin}
W. Rudin, \emph{Principles of Mathematical Analysis}, 3rd ed.
New York, NY, USA: McGraw--Hill, 1976.

\bibitem{danskin}
J. M. Danskin, \emph{The Theory of Max-Min and its Application
to Weapons Allocation Problems}.
Berlin, Germany: Springer-Verlag, 1967.
\end{thebibliography}
\end{document}